\documentclass{article}

\usepackage{arxiv}

\usepackage[utf8]{inputenc} 
\usepackage[T1]{fontenc}    
\usepackage{hyperref}       
\usepackage{url}            
\usepackage{booktabs}       
\usepackage{amsfonts}       
\usepackage{nicefrac}       
\usepackage{microtype}      
\usepackage{lipsum}		
\usepackage{graphicx}
\usepackage{natbib}
\usepackage{booktabs}
\usepackage{doi}
\usepackage{physics}
\usepackage{float}
\usepackage{subcaption}
\usepackage{multirow}
\usepackage{xcolor}
\usepackage{romannum}

\usepackage[authormarkup=none]{changes}
\definechangesauthor[name={Daniel Badawi}, color=red]{ec}

\title{Neural Operator-Based Proxy for Reservoir Simulations Considering Varying Well Settings, Locations, and Permeability Fields}

\author{
	Daniel Badawi\\
	Department of Petroleum Engineering\\
	Texas A\&M University\\
	daniel\_88@tamu.edu\\
	\\
	\And
	Eduardo Gildin\\
	Department of Petroleum Engineering\\
	Texas A\&M University\\
	egildin@tamu.edu\\
	\\
}

\date{}

\renewcommand{\shorttitle}{Neural Operator-Based Proxy for Two-Phase Flows in Porous Media}

\hypersetup{
    pdftitle={Neural Operator-Based Proxy for Reservoir Simulations Considering Varying Well Settings, Locations, and Permeability Fields},
    pdfsubject={},
    pdfauthor={Daniel Badawi, Eduardo Gildin},
    pdfkeywords={Neural Operators, Fourier Neural Operators, Transport in Porous Media, Reservoir Engineering, Reservoir Simulations, Darcy Flow},
}

\begin{document}
\pagenumbering{arabic}
\maketitle

\begin{abstract}
Simulating Darcy flows in porous media is fundamental to understand the future flow behavior of fluids in hydrocarbon and carbon storage reservoirs. Geological models of reservoirs are often associated with high uncertainly leading to many numerical simulations for history matching and production optimization. Machine learning models trained with simulation data can provide a faster alternative to traditional simulators. In this paper we present a single Fourier Neural Operator (FNO) surrogate that outperforms traditional reservoir simulators by the ability to predict pressures and saturations on varying permeability fields, well locations, well controls, and number of wells. The maximum-mean relative error of 95\% of pressure and saturation predictions is less than 5\%. This is achieved by employing a simple yet very effective data augmentation technique that reduces the dataset size by 75\% and reduces overfitting. Also, constructing the input tensor in a binary fashion enables predictions on unseen well locations, well controls, and number of wells. Such model can accelerate history matching and reservoir characterization procedures by several orders of magnitude. The ability to predict on new well locations, well controls, and number of wells enables highly efficient reservoir management and optimization.
\end{abstract}

\keywords{Neural Operators\and Fourier Neural Operators\and Transport in Porous Media\and Reservoir Engineering\and Reservoir Simulations\and Darcy Flow \and Machine Learning \and Deep Learning}

\section{Introduction}

\label{sec:intro}

Many machine learning techniques including deep neural networks have been recently used to simulate and predict flows in porous media. Such techniques can be completely data-driven or completely physics-informed with some utilizing both data and physics (hybrid). In general, data-driven deep learning has found applications variety of fields, including computer vision \cite{NIPS2012_c399862d}, natural language processing \cite{NLP}, speech recognition \cite{speech}, recommendation systems \cite{recommendation_systems}, autonomous vehicles \cite{autonomous}, healthcare (e.g., medical image analysis), and more. In our case, it found its way to reservoir simulations.

Data-driven deep models require large datasets to train them and usually don't extrapolate predictions outside the training parameter space \cite{ALMAJID2022109205}. Additionally, data-driven models are not committed to satisfy any physical laws that govern a dataset, thus leading to predictions that may be physically inconsistent or implausible \cite{Karniadakis}. Consequently, data-driven models or surrogates fall short to compete with traditional reservoir simulations because one of the main purposes of reservoir simulations is to predict future behavior of reservoir fluids and thus extrapolation in time is essential.

On the other hand, physics-informed deep learning models seek to embed the laws and principles of physics directly into deep neural network models. It acknowledges that physical systems follow fundamental laws, and these laws can be leveraged to improve the performance and reliability of deep learning models. Contrary to data-driven models, physics-informed models don't require training data, however, they can integrate data in their training process and such models are often referred to as hybrid models. Hybrid models are often used when there is a limited amount of data of a system with partly understood physics. The data compensate for the missing physics and enables inverse estimation of unknown physical properties.

Many data-driven models have been adopted in reservoir engineering for history matching, optimization, and production forecasting, \cite{Cao_production, Xiong_forcasting, Zhao_optimization, Ma_, Tian_}. For physics-based models, \cite{raissi} introduced the famous physics-informed neural network (PINN), a general framework of using neural networks (NNs) to solve PDEs in fluid mechanics and engineering. Following the idea of PINNs, many models have been introduced to solve PDEs involved in reservoir engineering. \cite{Fraces, emilio_, originai, jingjing} used physics-informed models to solve the Buckley-Everett equation. \cite{badawi2023physics}, used the spatial domain decomposition to solve the transient radial diffusivity equation, additionally, leveraging the inverse capabilities of PINNs with flow rates data \cite{badawi2023physics} estimated missing physical properties such as permeability and distance from fault. \cite{ZHANG2022110179, ZHANG2023111919}, proposed the physics-informed convolutional neural network (PICNN) to build models to simulate flow in porous media over a grid, and surrogates to predict pressures and saturations on unseen permeability fields. PICNNs models appear to be able to simulate flow in porous media, however they are highly constrained by very small time-steps to converge. PICNNs surrogates although gave good results in \cite{ZHANG2022110179, ZHANG2023111919}, they appear to lose quality and accuracy when applied to reservoirs with multiple wells and complicated well control schedules. Others proposed models that incorporate simulated data with PINNs to obtain surrogates that can simulate flows in porous media \cite{tartakovsky2020physics}, \cite{WANG2021113492}, \cite{WU2021100044}).

Reservoir simulations are very computationally expensive mainly due to the large scale of reservoir models. A typical reservoir model size ranges between hundred thousand up to tens of millions of grid-blocks and therefore a single simulation run can extend up to ten hours or more on a supercomputer. A single reservoir characterization and history matching job needs hundreds up to thousands of such simulations. Additionally, the boundary conditions (BCs) associated with wells are constantly changing in time. Some of the well BCs include wells shut-in, drilling new wells, and change in well controls (bottom-hole pressure (BHP) or rates). Data-driven models usually don't extrapolate outside the training domain, and physics-informed neural network models work only with fully defined physical problems and can not adapt to changing boundary conditions such as well location and well controls or changing physical properties such as permeability and porosity fields. This is where neural operators become handy.

Neural operators can learn and generalize from data, enabling fast predictions once trained. They provide a more flexible and scalable approach, capable of handling varying boundary conditions and geometries. Moreover, neural operators can improve accuracy and reduce computational costs by leveraging advanced machine learning techniques, making them highly suitable for real-time applications. This perfectly aligns with our interest in building reservoir surrogate that predicts on varying boundary conditions mentioned above.

In this paper we utilize Fourier neural operator (FNO) to build two surrogates; the first one is suitable for single phase flows which provides predictions of pressure, and the second one provides predictions of both pressures and saturations. Both surrogates predict on new permeability fields, new well locations, new well controls, and different number of wells, as well as forward predictions in time.
\section{Neural Operators}
\label{sec:NO}
A neural operator is a data-driven approach that aims to directly learn the solution operator of PDEs from data. Unlike neural networks that learn function mapping between finite-dimensional spaces, neural operator extends that to learning the operator between infinite-dimensional spaces \cite{DeepOnets}, \cite{MOD_NeuralNets}, \cite{NO_3}, \cite{PATEL}, \cite{Li2020NeuralOG}. This approach enables learning solutions of a wide set of parameterized PDEs with wide in a variety of BCs. This aligns with our interest in building a surrogate that accurately predicts pressure on new permeability fields as well as new well locations, controls, and number of wells. In particular we are interested in the application of Fourier Neural Operator (FNO) because it has been shown that it outperforms some other neural operators \cite{Li2020FourierNO}. An interesting feature that automatically results from the formulation of the FNO architecture is the discretization invariance, meaning that the model can be trained on coarse grid samples and evaluated (tested) on finer grid samples. This can be a very important feature when dealing with large-scale reservoir models. In this paper, this feature is not going to be utilized or tested, we are just mentioning it for the knowledge of the reader.

The neural operator proposed by \cite{Li2020NeuralOG}, is an iterative procedure $v_0 \mapsto v_1,...,v_T$ where $v_j$ for $j=1,2,...T-1$ is a series of functions taking values in $\mathbb{R}^{d_v}$. As shown in Fig. \ref{fig:ufno}(a), the input $a(x) \in \mathcal{A}(D;\mathbb{R}^{d_a})$ is lifted to higher dimensional channel space $v_0(x)=P(a(x))$ by the local transformation $P$ which is parameterized by a simple fully-connected neural network. Then several iterative updates $v_t \mapsto v_{t+1}$ are applied. Each update $v_t \mapsto v_{t+1}$ is composed of a non-local integral operator $\mathcal{K}$ and a local nonlinear activation function $\sigma$. The output $u(x) \in \mathcal{U}(D;\mathbb{R}^{d_u})$ is obtained by projecting $v_T$ by a local transformation $Q:\mathbb{R}^{d_v} \rightarrow \mathbb{R}^{d_u}$, that is, $u(x)=Q(v_T(x))$.
The input space $\mathcal{A}=\mathcal{A} (D;\mathbb{R}^{d_a})$ and output space $\mathcal{U}=\mathcal{U}(D;\mathbb{R}^{d_u})$ are separable Banach spaces of functions taking values in $\mathbb{R}^{d_a}$ and $\mathbb{R}^{d_u}$ respectively.

The iterative update $v_t \mapsto v_{t+1}$ is defined as follows:
\begin{align}\label{iter_update}
    v_{t+1}:=\sigma\bigg( Wv_t+\big(\mathcal{K}(a;\phi)v_t\big)(x)\bigg); \hspace{1cm} \forall x \in D; \hspace{0.5cm} D \subset \mathbb{R}^d
\end{align}
where $\mathcal{K} : \mathcal{A} \cross \Theta_{\mathcal{K}} \rightarrow \mathcal{L}\big(\mathcal{U}(D;\mathbb{R}^{d_v}), \mathcal{U}(D;\mathbb{R}^{d_v})\big)$ maps to bounded linear operators on $\mathcal{U}(D;\mathbb{R}^{d_v})$ and is parameterized by $\phi \in \Theta_{\mathcal{K}}$, $W:\mathbb{R}^{d_v} \rightarrow \mathbb{R}^{d_v}$ is a linear transformation, and $\sigma:\mathbb{R} \rightarrow \mathbb{R}$ is a non-linear activation function whose action is defined point-wise.

The kernel integral operator $\mathcal{K}$ in (\ref{iter_update}) is defined by
\begin{align}\label{kernel_int_op}
    \mathcal{K}=\int_D\kappa\big(x,y,a(x),a(y);\phi\big)v_t(y)dy; \hspace{1cm} \forall x \in D; \hspace{0.5cm} D \subset \mathbb{R}^d
\end{align}
where $\kappa_\phi: \mathbb{R}^{2(d+d_a)} \rightarrow \mathbb{R}^{d_v \cross d_v}$ is a neural network parameterized by $\phi \in \Theta_\mathcal{K}$. Now $\kappa_\phi$ replaces the kernel function $\kappa$ in (\ref{kernel_int_op}) which we can learn from data. 

Moreover, disconnecting the dependence of  $\kappa_\phi$ from the input function $a$ and imposing $\kappa_\phi(x; y) =\kappa_\phi(x-y)$, we obtain that (\ref{kernel_int_op}) is a convolution operator. Exploiting this fact by parameterizing $\kappa_\phi$ directly in Fourier space and using the Fast Fourier Transform (FFT) to efficiently compute (\ref{kernel_int_op}) leads to the Fourier neural operator architecture.

\subsection{Fourier Neural Operator}
In this section we will not discuss the derivation of FNO, instead, we will briefly address the architecture assembly. For more details on the mathematical derivation refer to \cite{Li2020FourierNO}.

\begin{figure}
    \centering
    \includegraphics[scale=0.38]{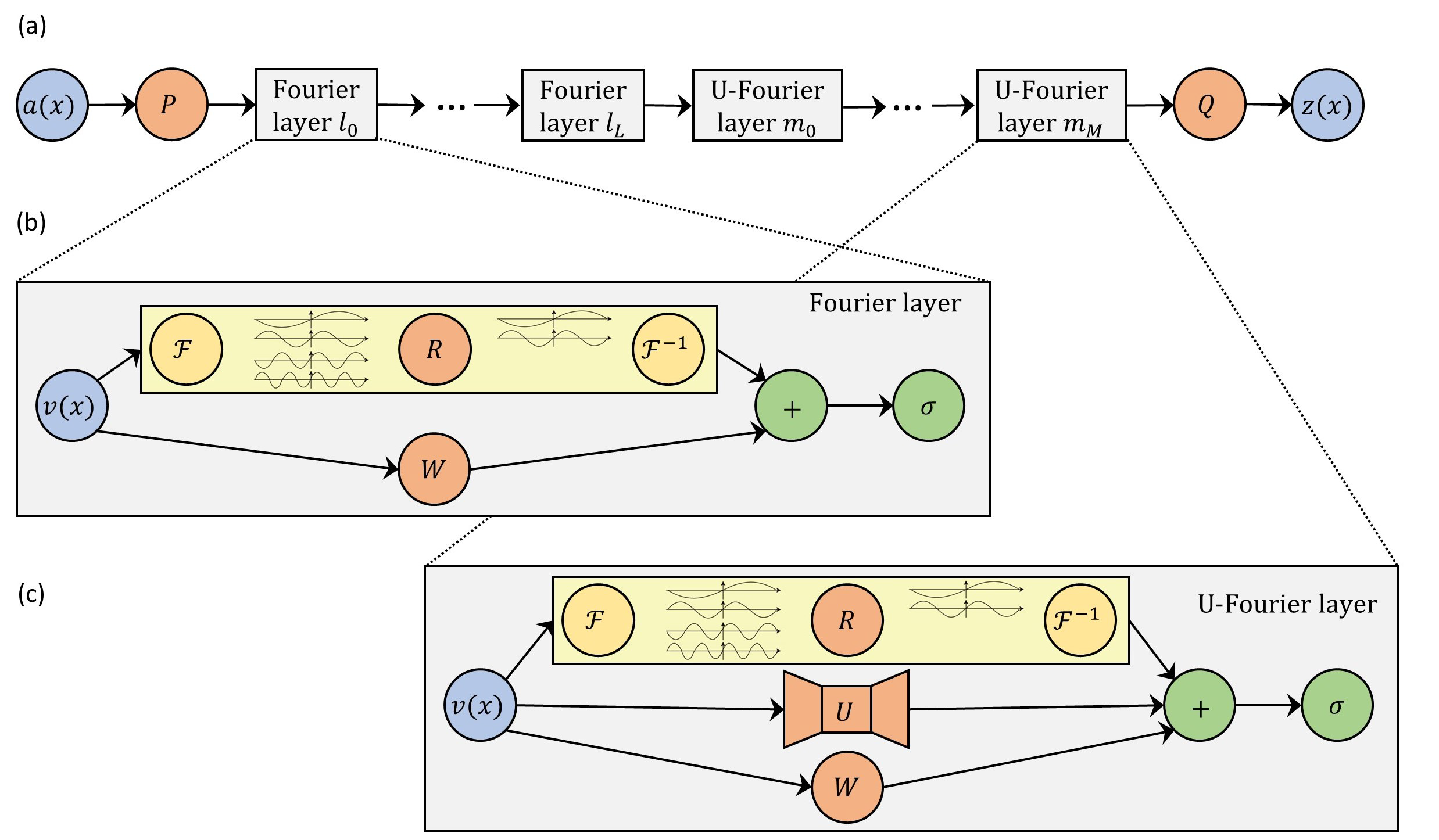}
    \caption{U-FNO architecture.}
    \label{fig:ufno}
\end{figure}

As illustrated in Fig. \ref{fig:ufno}a, a general FNO model constitutes of a lifting layer $P$, multiple Fourier blocks, and a down-sampling layer $Q$. The lifting layer $P$ is a fully connected network that lifts the physical input $a(x)$ to higher dimension channel space. After that, the output $v(x)$ passes through a sequence of Fourier blocks. The output from the last Fourier block is fed to $Q$ which is also a fully connected network (or more) that decodes the output in the high dimensional channel space back to the physical space. $Q$ performs the opposite operation of $P$.

A Fourier block constitutes of a Fourier transform layer $\mathcal{F}$ that transforms the input $v(x)$ into Fourier space followed by a linear transform $R$ (matrix multiplication) applied to the lower Fourier modes and then an inverse Fourier layer $\mathcal{F}^{-1}$ that projects the output from Fourier space back to the high dimension space real space. $R$ is essentially a trainable parameters tensor that multiplies the low Fourier modes. In this work the number of Fourier modes is 10. In parallel, the input $v(x)$ in passed through a local linear transform $W$ (bias term) which is typically a simple convolution layer. Since Fourier transforms work only with periodic BCs, the bias term $W$ is essential when working with non-periodic boundary conditions as it retains record of such BCs. The outputs from $W$ and $\mathcal{F}^{-1}$ are summed and fed into a non-linear function $\sigma$ which is ReLU in our case. A schematic of a Fourier block is displayed in Fig. \ref{fig:ufno}b.

In this work we use the enhanced FNO namely U-FNO which has proved to provide better performance for similar cases to ours \cite{ufno}. The difference between regular FNO and U-FNO is that in the U-FNO architecture we add a parallel U-net inside some of the last Fourier blocks as demonstrated in Fig. \ref{fig:ufno}c. Fourier blocks with a U-net are labeled as U-Fourier blocks. In this work the number of U-Fourier blocks is 3.

\section{Model Problem}
Two-phase oil-water flow in porous media is expressed by a set of partial differential equations which can be derived by combining the material balance equation and Darcy's law. Mathematically, it is expressed as below: 

\begin{align}\label{diff_eq_2ph}
    \frac{\partial\big(\phi \rho_j S_j\big)}{\partial t} - \nabla\bigg[\rho_i \lambda_j \mathbf{K}_j\big(\nabla p_j-\rho_j g \nabla D\big)\bigg] + q_j^{ss} = 0
\end{align}

where subscript $j$ corresponds to the phase ($w$-water, $o$-oil). $\rho_j$ is phase density, $\phi$ is porosity, $\mathbf{K}_j$ is phase permeability tensor, $\lambda_j$ is phase mobility $\lambda_j=k_{rj}/\mu_j$ with $k_{rj}$ and $\mu_j$ are phase relative permeability and phase viscosity respectively. $g$ is gravitational acceleration, $D$ is depth, and $q_j^{ss}$ is the phase sources/sinks term. $p_j$ and $S_i$ are phase pressure and saturation respectively. It is important to note that eq. \ref{diff_eq_2ph} is a two equation set with four unknowns being $(p_w, p_o, S_o, S_w)$. This is completed by introducing two additional equations; the saturation constraint $S_w+S_o=1$, and the capillary pressure relationship $p_c(S_w)=p_o-p_w$. Eq. \ref{diff_eq_2ph} is a nonlinear set of PDEs due to the dependency of rock and fluid parameters on pressure and saturation. For single-phase flow eq. \ref{diff_eq_2ph} reduces to the following

\begin{align}\label{diff_eq_1ph}
    \frac{\partial\big(\phi \rho\big)}{\partial t} - \nabla\bigg[\frac{\rho \mathbf{k}}{\mu}\big(\nabla p-\rho g \nabla D\big)\bigg] + q^{ss} = 0
\end{align}

where $\mathbf{k}$ being the absolute permeability tensor. In this work the gravity terms $\big(\rho g \nabla D\big)$ in eqs. \ref{diff_eq_2ph} and \ref{diff_eq_1ph} are neglected.

\section{FNO Implementation}
\label{ufno_implement}
In this paper we present two FNO models; the first is compatible to single-phase flows that predicts pressure distribution; the second is compatible to two-phase flows that predicts both pressures and saturations. Both models predict on new permeability fields, well locations, number of wells, and well controls. FNOs are data-driven models that acknowledge that a dataset is governed by physical laws and learns these laws by minimizing an appropriate loss function. 

For training the single-phase model, we generated a dataset containing 1400 permeability fields using Stanford Geostatistical Modeling Software (SGeMS) \cite{sgems}, with 1000 samples for training and 400 for validation. For testing we used the same 400 permeability fields used for validation with different well controls, well locations and number of wells. The spatio-temporal size of each sample is $(N_x,N_y,N_t)=(40,40,40)$. The simulation time is range is $[0,440]$ days. The time-step size between two consequent snapshots $\Delta t=2$ days, i.e., ($N_{ss}=221$). All training and validation samples have 6 wells randomly distributed in space with four being production wells and two injection wells, whereas, testing samples include different number of wells ranging from a single well up to nine wells. This means that the model is trained and validated on samples containing 6 wells only, and tested on samples containing up to 9 wells. Moreover, every well has a unique control settings generated using random uniform distributions. Finally, to test forward extrapolation in time, the training time range is constrained to $[0,280]$ days, and validation-testing time range is $[280,440]$ days. In terms of snapshots $N_{t_{tr}} \in [0,140]$ and $T_{vl},T_{ts} \in [140,221]$.

For training the two-phase model, we generated simulations for 4200 samples, with 3500 samples for training, 500 for validation and 200 for testing. The simulation time is range is $[0,10]$ years with time-step size between two consequent snapshots $\Delta t=10$ days, i.e., ($N_{ss}=366$). Similar to the first model all training and validation samples have 6 wells randomly distributed in space with four being production wells and two injection wells, whereas, testing samples include different number of wells ranging from three wells up to twelve wells. Training time range is constrained to $[0,2380]$ days, and validation-testing time range is $[0,3650]$ days. Similar to the first model all well controls are generated using random uniform distributions with no practical or engineering considerations.

It is important to note that in the first model an injector is defined as a well with BHP maintained above the initial pressure $p_0$ which is $30 [MPa]$, and the opposite for a producer. For the second model producers are controlled by bottom-hole pressure and injectors are controlled by injection rates. All simulations were obtained using CMG IMEX simulator.

\subsection{\textbf{Inputs and Outputs}}
\begin{figure}
    \centering
    \includegraphics[scale=0.28]{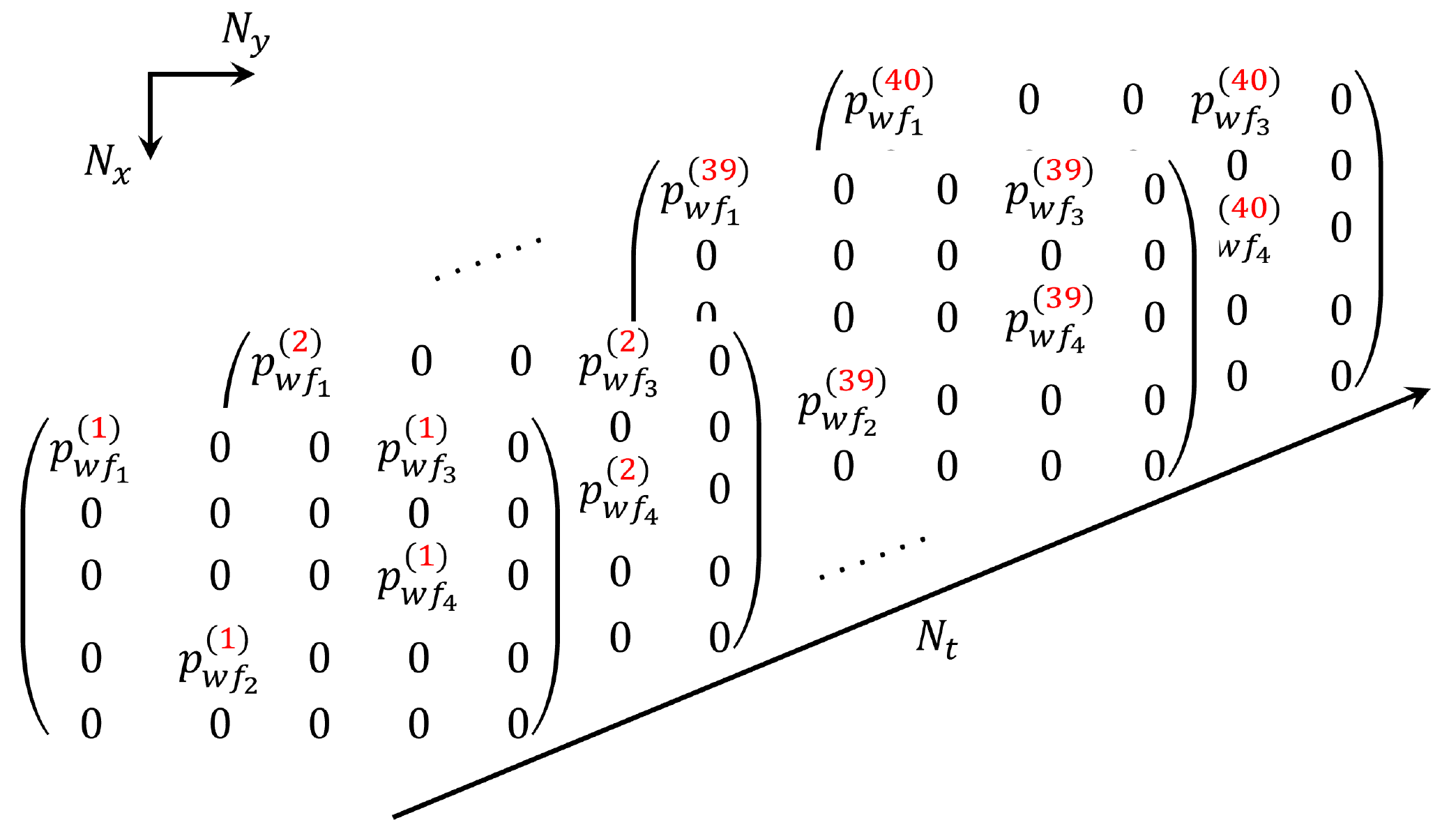}
    \caption{Pressure Well controls (BHP) for a single $(5x5x40)$ sample for demonstration purposes only. }
    \label{fig:well_ctrls_channel}
\end{figure}

The input to the FNO is 5D tensor $(N_s, N_c, N_x, N_y, N_t)$, with $N_s$ being the number of samples, $N_x=40$, $N_y=40$, $N_t=40$ the number of discretization in space and time, and $N_c$ number of channels which is 4 for the single-phase model and 6 for the two-phase model. The channels $N_c$ for the single-phase model are permeability, well controls (BHPs), well locations, and pressure initial condition. For the two-phase model the channels are permeability, producer controls (BHPs), injector controls (rates), well locations, pressure initial condition, and saturation initial condition. For a single sample, the well location channel is a $(40, 40, 40)$ binary tensor with 1 where is an active well and zeros elsewhere. The well controls channel similar to the well locations channel with the actual values of the well controls as in Fig. \ref{fig:well_ctrls_channel}.

The output shape is $(N_s, N_c, N_x, N_y, N_t)$, where $N_c=1$ is the predicted pressure for the single-phase model and $N_c=2$ predicted pressure and saturation for the two-phase model.

\subsection{\textbf{Loss Function}}
\begin{figure}
    \centering
    \includegraphics[scale=0.7]{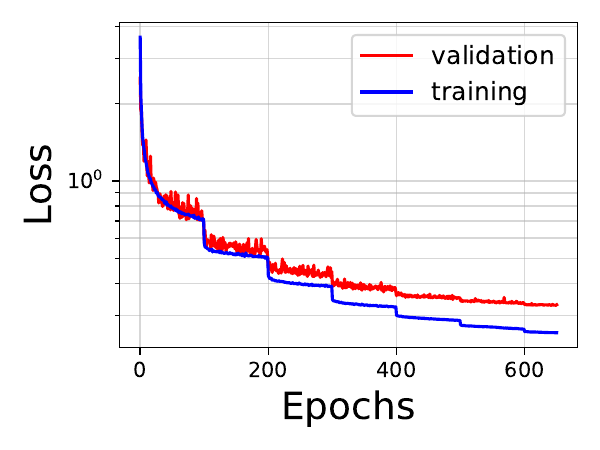}
    \caption{Loss function for the two-phase model.}
    \label{fig:loss_2ph}
\end{figure}

Training is performed by minimizing the relative L2-loss function which has proved to be more efficient than the mean-square error (MSE). Also, the relative L2-loss of pressure and saturation derivatives are incorporated in the total loss. Adding the derivative component has shown to accelerate convergence. The loss function equation for the single-phase and two-phase models are given in Eqs. \ref{loss1ph}, \ref{loss2ph} respectively.
\begin{align}\label{loss1ph}
    \mathcal{L}\big(p,\hat{p})\big) = L_2(p,\hat{p}) + \alpha\sqrt{L_2\bigg(\frac{\partial p}{\partial x}, \frac{\partial \hat{p}}{\partial x}\bigg)^2 + L_2\bigg(\frac{\partial p}{\partial y}, \frac{\partial \hat{p}}{\partial y}\bigg)^2}
\end{align}

\begin{equation}\label{loss2ph}
\begin{split}
    \mathcal{L}\big([p,S_w],[\hat{p},\hat{S_w}]\big) = L_2\big([p,S_w],[\hat{p},\hat{S_w}]\big) + \alpha_1\sqrt{L_2\bigg(\frac{\partial p}{\partial x}, \frac{\partial \hat{p}}{\partial x}\bigg)^2 + L_2\bigg(\frac{\partial p}{\partial y}, \frac{\partial \hat{p}}{\partial y}\bigg)^2} + &\\ + \alpha_2\sqrt{L_2\bigg(\frac{\partial S_w}{\partial x}, \frac{\partial \hat{S_w}}{\partial x}\bigg)^2 + L_2\bigg(\frac{\partial S_w}{\partial y}, \frac{\partial \hat{S_w}}{\partial y}\bigg)^2}
\end{split}
\end{equation}

where $p$ is the true pressure, $\hat p$ is the predicted pressure, $S_w$ is the true saturation, $\hat S_w$ is the predicted saturation, and $\alpha$'s are weighting factors equal to 1.0 for both models. $L_2$ is given by
\begin{align}
    L_2(z,\hat{z}) = \frac{\norm{z - \hat{z}}_2}{\norm{z}_2}
\end{align}
Initially, we tried a different form of loss function given in Eqs. \ref{loss1}, \ref{loss2}, however it didn't produce satisfactory results.

\begin{align}\label{loss1}
    \mathcal{L}(p,\hat{p}) = L_2(p,\hat{p}) + \alpha \bigg[L_2\bigg(\frac{\partial p}{\partial x}, \frac{\partial \hat{p}}{\partial x}\bigg) + L_2\bigg(\frac{\partial p}{\partial y}, \frac{\partial \hat{p}}{\partial y}\bigg)\bigg]
\end{align}
\begin{align}\label{loss2}
\begin{split}
    \mathcal{L}\big([p,S_w],[\hat{p},\hat{S_w}]\big) = L_2\big([p,S_w],[\hat{p},\hat{S_w}]\big) + \alpha_1 \bigg[L_2\bigg(\frac{\partial p}{\partial x}, \frac{\partial \hat{p}}{\partial x}\bigg) + L_2\bigg(\frac{\partial p}{\partial y}, \frac{\partial \hat{p}}{\partial y}\bigg)\bigg] + &\\ + \alpha_2 \bigg[L_2\bigg(\frac{\partial S_w}{\partial x}, \frac{\partial \hat{S_w}}{\partial x}\bigg) + L_2\bigg(\frac{\partial S_w}{\partial y}, \frac{\partial \hat{S_w}}{\partial y}\bigg)\bigg]
\end{split}
\end{align}
The loss function of the two-phase model is shown in Fig. \ref{fig:loss_2ph}.
\subsection{\textbf{Masking}}
The saturation plume speed travels much slower than the pressure wave, therefore grid-blocks where the water plume hasn't reached experience no change in their initial saturation $S_{w0}$, hence we mask these grid-blocks so the learning process focuses only on grid-blocks that already experienced change in their initial saturation. We realize that this practice accelerates convergence. Note that masking is performed for saturations only, i.e., pressure and derivatives are not masked.

\subsection{\textbf{Scaling and Normalization}}
In general, scaling is a very recommended preprocessing step in machine learning, especially when input channels has significant difference in their order of magnitude. For example, permeability usually has an order of magnitude between $[10^{-18},10^{-13}]$ $m^2$, whereas pressures usually have order of magnitude of $\sim10^6$ $Pa$. Therefore, the inputs must be scaled between $[0,1]$. For pressure and rates attributed channels (well controls, initial conditions, and outputs), we simply divide by the maximum value of the controls respectfully, and for permeability we scale using the logarithmic min/max scalar as below
\begin{align}
    p_{sc}=\frac{p}{p_{max}^{ctrl}}; \hspace{0.5cm} q_{sc}=\frac{q}{q_{max}^{ctrl}}
\end{align}
\begin{align}
    K_{sc}=\frac{log(K)-log(K_{min})}{log(K_{max})-log(K_{min})}=\frac{log(K/K_{min})}{log(K_{max}/K_{min})}
\end{align}

\subsection{\textbf{Data Augmentation}}
Reservoir simulations at large scale are computationally expensive, and since our FNO models are data-driven models, a large dataset (simulations) is required for training. To increase the size of the training set with minimal simulations and to enrich the training set, we used data augmentation by flipping each sample to the right, down and down-right. In general, data augmentation by flipping is widely used in various image classification CNN models. This helps the model generalize better to unseen data and reduces the risk of overfitting, \cite{augmentation}.  

In this section, we introduce a data augmentation concept and without loss of generality, we are showing the augmentation results for the single-phase model, however, the concept is also utilized in training the two-phase model.

Although our models are not classification models, we know that simulating a flipped input sample produces a flipped pressure field, and thus a flipped input sample can be a new sample without the need to simulate the original sample. To study the effects of the augmentation scheme we are proposing, we trained four distinct single-phase models on four different training sets. The first training set contains the 250 base samples (simulations), the second training set contains the same 250 base samples of the first set with augmentation, thus, the size of the second training set is 1000. The third training set contains 1000 base samples, and the fourth training set contains the same 1000 base samples of the third set with augmentation and thus the size of the fourth training set is 4000 samples.

All models are trained with batch size of 10, therefore, to train all models the same amount of batch passes (same number of parameter updates), model 1 was trained with 9600 epochs, model 2 and 3 were trained with 2400 epochs, and model 4 was trained with 600 epochs. This guarantees that the number of parameter updates for all models is the same which is 240,000 updates (batch passes) in our case.

\begin{figure}
    \centering
    \includegraphics[scale=0.35]{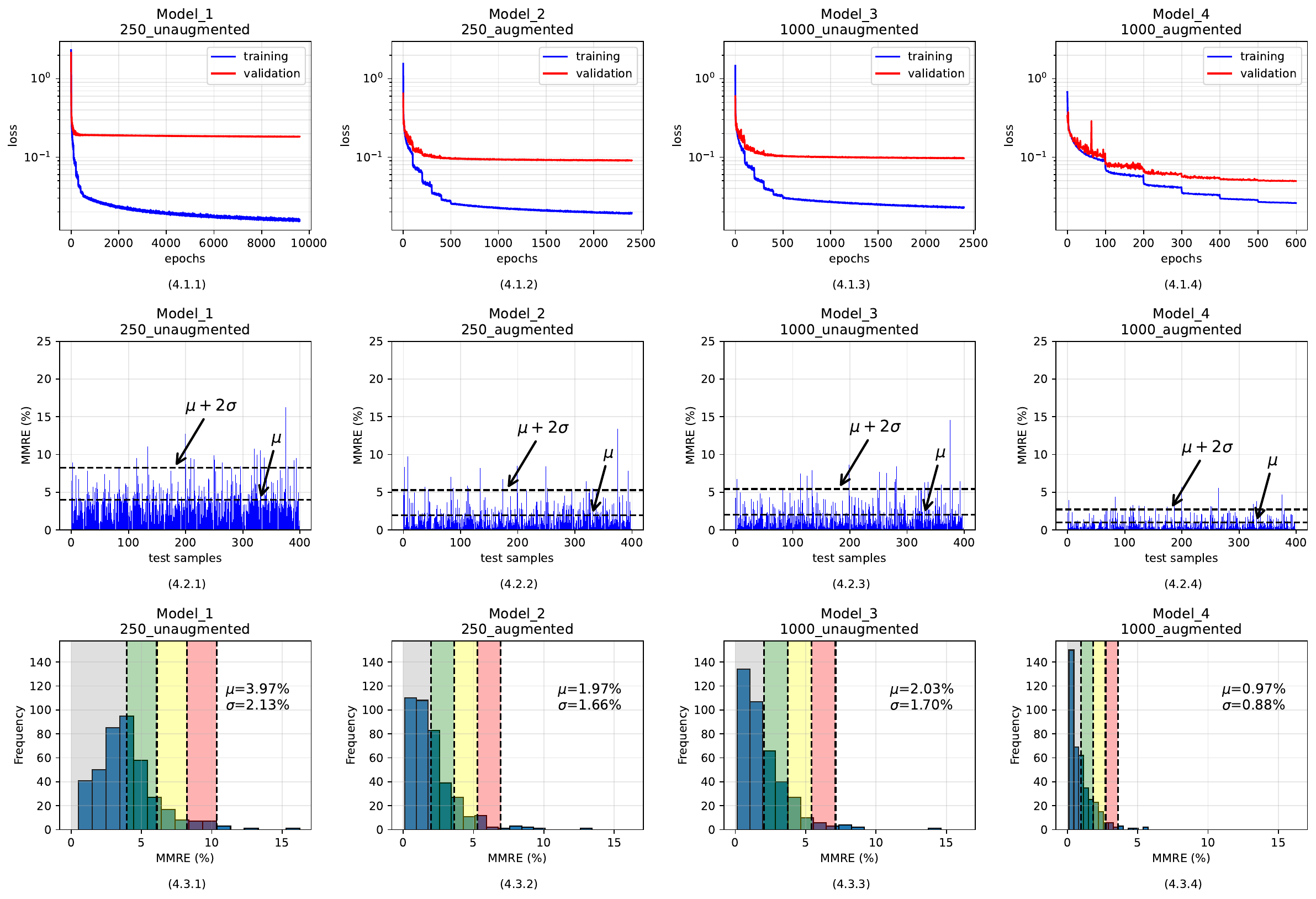}
    \caption{Data augmentation. Column (1) illustrates the statistical analysis of the first model trained with 250 \textbf{unaugmented} samples dataset. Column (2) illustrates the statistical analysis of the second model trained with 250 \textbf{augmented} samples dataset. Column (3) illustrates the statistical analysis of the third model trained with 1000 \textbf{unaugmented} samples dataset. Column (4) illustrates the statistical analysis of the fourth model trained with 1000 \textbf{augmented} samples dataset. In the second row, the two dashed lines represent the mean and mean plus two standard deviations respectively, and in the third row, the gray, green, yellow, red areas are the mean, mean plus one, two, and three standard deviation respectfully.}
    \label{fig:aug}
\end{figure}

It is evident from Fig. \ref{fig:aug} that data augmentation indeed enhanced the results. The first row shows the training and validation loss for the four models. The second row shows the maximum mean relative error MMRE (given by Eq. \ref{MMRE}) for all 400 validation samples for the four models. The third row is the histogram of the MMRE of the four models. Note that the validation set is the same for all four models.

For the first model, the validation loss flattens very quickly (Fig. \ref{fig:aug}.1.1), with training loss being the lowest among all models. This suggests that the model is highly overfitting. This is quite expected since 250 samples for training are not sufficient for generalizing. The second model which is trained on the same 250 base samples of model 1 with augmentation clearly shows significant improvement. This is firstly seen in Fig.\ref{fig:aug}.1.2 as the gap between the training and validation loss (overfitting) is reduced, and secondly in Fig.\ref{fig:aug}.2.2 and Fig.\ref{fig:aug}.3.2 as the mean $(\mu)$ and standard deviation $(\sigma)$ of MMRE are reduced from 3.97\% and 2.13\% in model 1 to 1.97\% and 1.66\% in model 2 respectively.

Moreover, from the results of model 2 and 3, it is evident that they are very comparable to each other, this is illustrated in Fig. \ref{fig:aug} column 2 and 3. In Fig. \ref{fig:aug}.1.2 and \ref{fig:aug}.1.3 we see that the overfitting is significantly reduced compared to model 1, however it still exists, and this is seen in the gap between the validation and training loss. In Fig. \ref{fig:aug}.3.2 and Fig. \ref{fig:aug}.3.3 the mean and standard deviation of the MMRE of both models are very close. To conclude, model 2 which was trained using 250 base samples with augmentation generalizes almost as equally as model 3 which was trained with 1000 base samples, and therefore, training model with augmentation reduces the number of training simulations by 75\%.

Regarding model 4 which is trained using 1000 base samples with augmentation (4000 training samples total), although it still exhibits some overfitting it is much less than model 3, and again this validates that augmentation indeed reduces overfitting. In Fig.\ref{fig:aug}.3.4, we see that model 4 generalizes more than model 3 with $\mu=0.97\%$ and $\sigma=0.88\%$. Overall, this analysis verifies that augmentation in general can be applied to similar problems. On one hand it reduces overfitting and enables better generalization, and on the other it reduces the number of training simulations.

\subsection{\textbf{Model Architecture and Training Parameters}}
\label{sec:architecture}
Table. \ref{fno_table} summarizes the model architecture. For the first model prediction (output) shape is $(1,40,40,40)$ corresponding pressure, and for the second model the prediction shape is $(2,40,40,40)$ corresponding to pressure and saturation. In this work the number of Fourier modes is 10 $(m=10)$. 

Note, the number of Fourier modes can not be larger than $N//2+1$, where $N=min\{N_x, N_y, N_t\}$, and training was done using Adam optimizer with step learning rate schedule that drops in half every 100 epochs.

\section{Results}
\label{sec:results}
\begin{figure}
    \centering
    \includegraphics[scale=0.23]{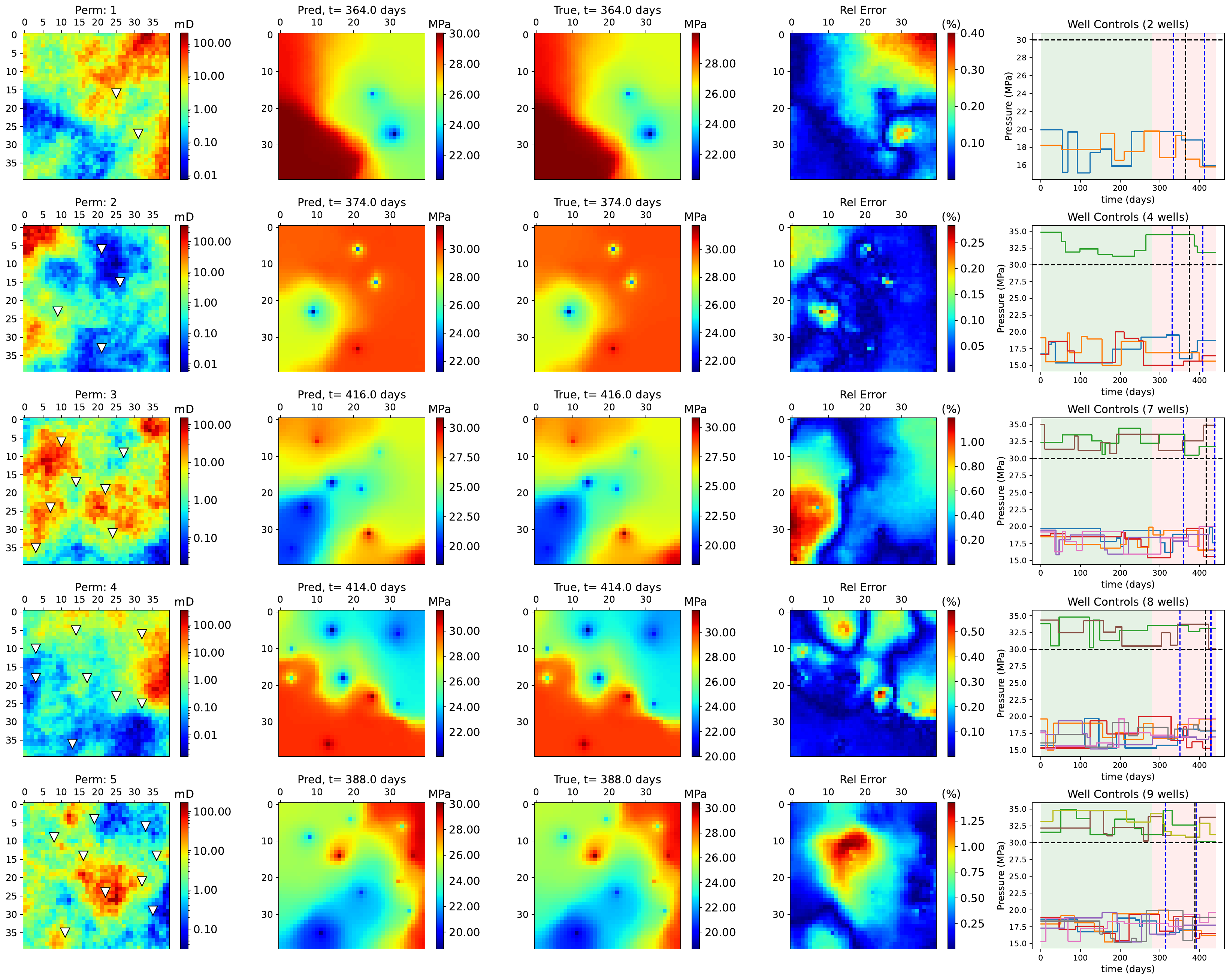}
    \caption{Model 1: pressure predictions for five testing samples over testing time range $[240, 440]$ days (red area) with new permeability fields, well locations, well controls, and number of wells. The first column is the permeability fields with the location of wells marked with white triangles on the same plot. The second column displays pressure predictions, the third column displays simulation pressure, the fourth column is the point-wise relative error, and the fifth column is the well controls for the testing time range. The horizontal black dash-line in the well controls plots is the initial pressure at $t=0$ days. The green and red areas represent the training and testing time range respectively. The two vertical dashed blue lines mark the 40 time-steps range of the test samples, and the vertical black dashline represents the timestep whose results are shown in the figure.}
    \label{fig:1ph_p_pred}
\end{figure}

\begin{figure}
    \begin{subfigure}{1.0\textwidth}
    \centering
    \includegraphics[scale=0.25]{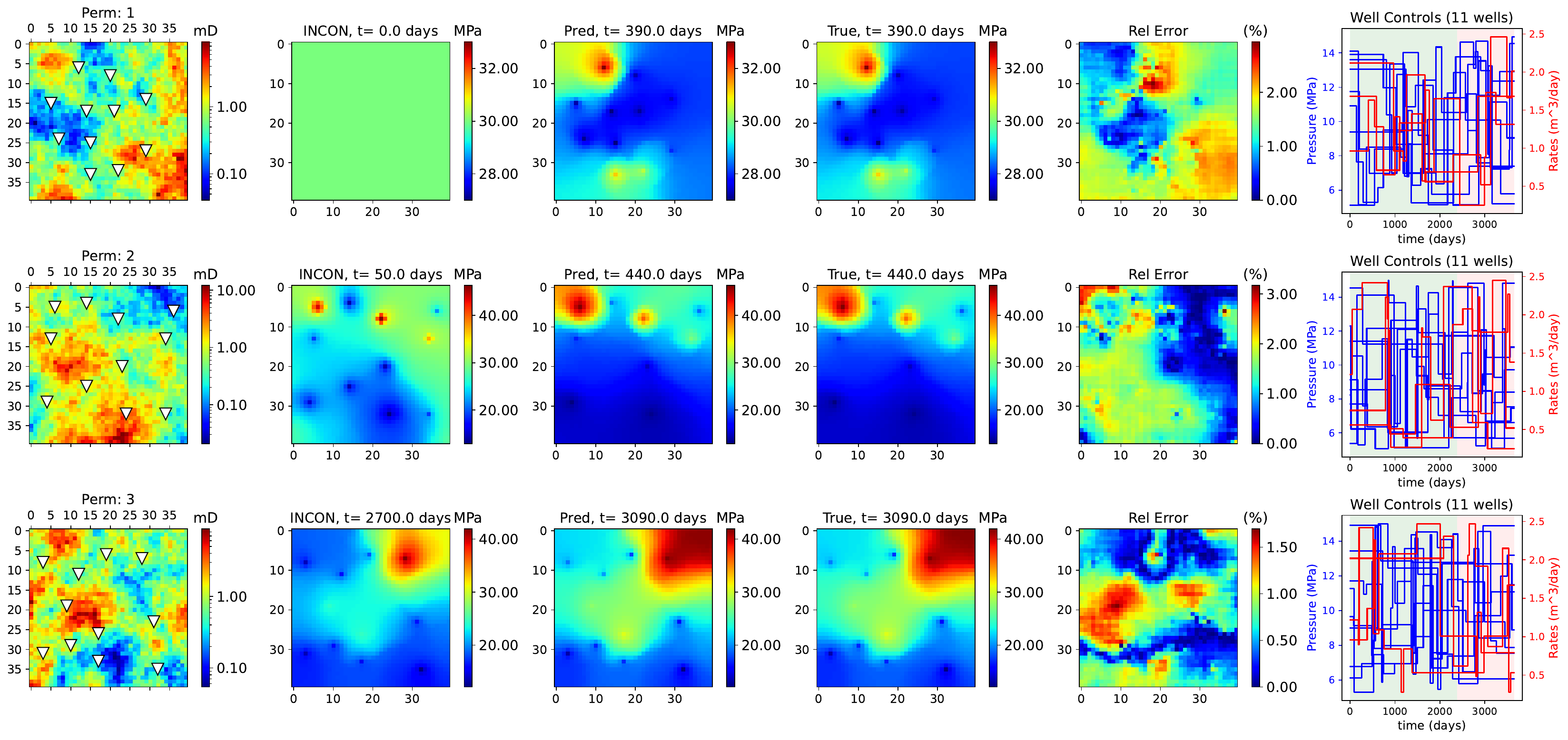}
    \caption{Pressure Predictions}
    \label{fig:2ph_p_pred}
    \end{subfigure}

    \begin{subfigure}{1.0\textwidth}
    \centering
    \includegraphics[scale=0.25]{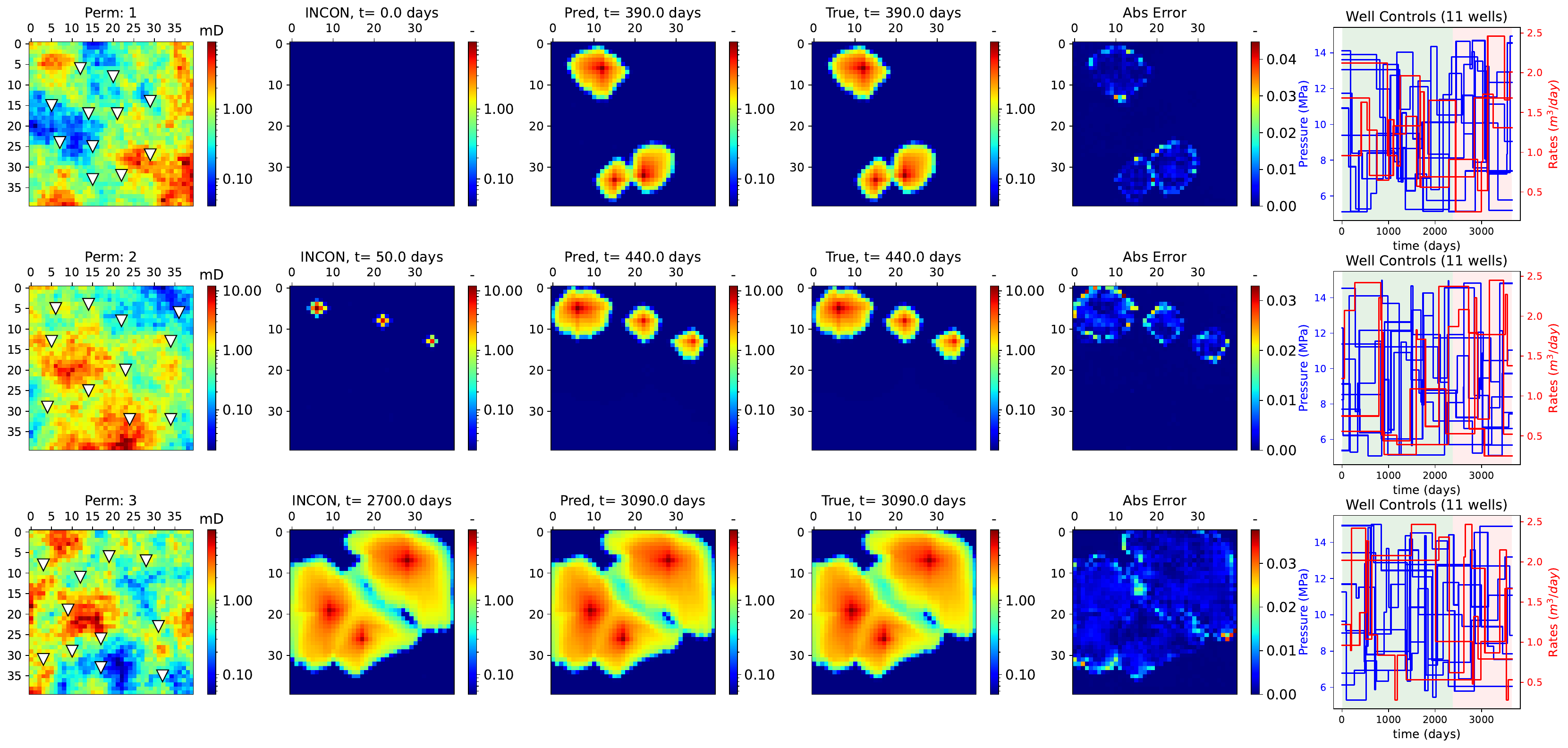}
    \caption{Saturation Predictions}
    \label{fig:2ph_sw_pred}
    \end{subfigure}
\caption{(a) - pressure predictions (b) - saturation predictions for three testing samples with new permeability fields, well locations, well controls, and number of wells. First column is permeability fields with the location of wells marked with white triangles on the same plot, second column is initial condition, third column is prediction, fourth column is true value, fifth column is the point-wise errors (relative/absolute), and sixth column is well controls.}
\end{figure}

In this section we present the results of the predictions for both models. Fig. \ref{fig:1ph_p_pred} displays predictions for five testing samples with unseen permeability fields, well locations, number of wells, well controls and different time range for the single-phase model. Similarly, Fig.\ref{fig:2ph_p_pred} and Fig.\ref{fig:2ph_sw_pred} show predictions of pressures and saturations of three samples respectively associated with the two-phase model.

Keep in mind that for the single-phase model training samples were taken in time range $[0,238]$ days, while the predictions are in future time range $[240,440]$ days. And for the two-phase model, training samples were taken in time range $[0,2380]$ days, while predictions are in time range $[0,3650]$ days.

\begin{figure}
\begin{minipage}{.45\textwidth}\label{stats1}
    \subfloat[]{\includegraphics[width=\textwidth]{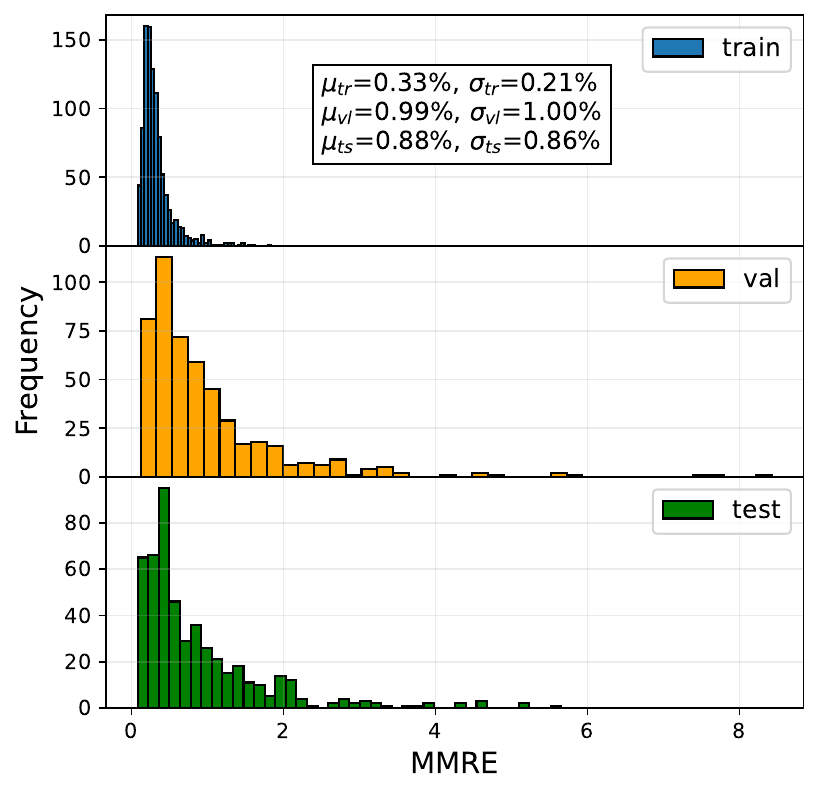}\label{fig:MMRE_1ph}}
\end{minipage}
\hfill    
\begin{minipage}{.45\textwidth}\label{stats2}
    \subfloat[]{\includegraphics[width=\textwidth]{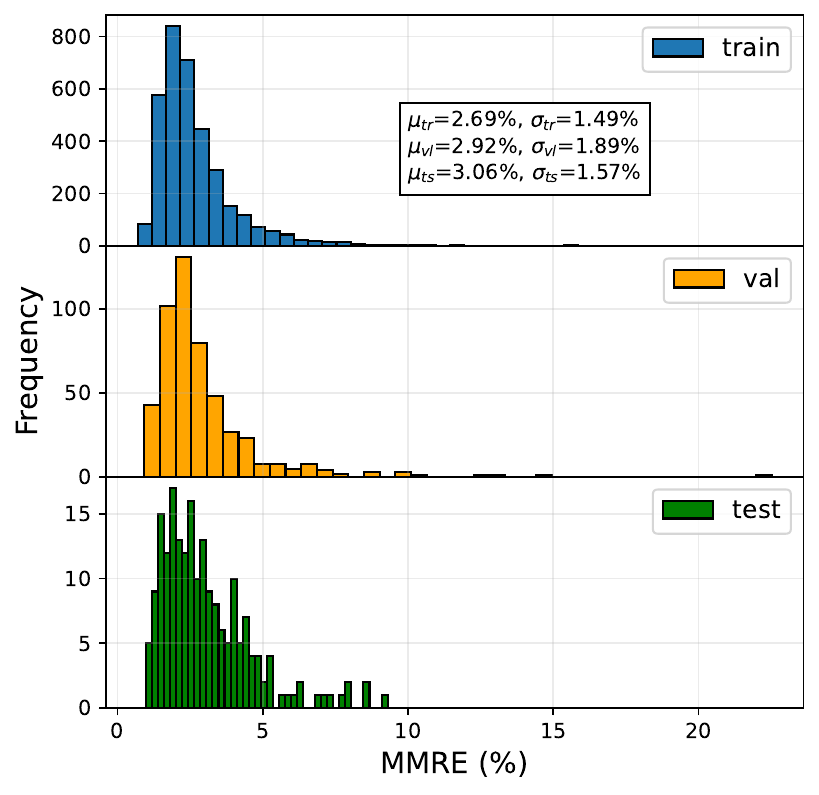}\label{fig:MMRE_2ph}}
\end{minipage}
    \caption{(a) - Histogram plot for pressure MMRE for the single-phase model with (1000, 400, 400) training, validation, testing split. (b) - Histogram plot for pressure MMRE for the two-phase model with (3500, 500, 200) training, validation, testing split.}\label{fig:stats}
\end{figure}

The ability to extrapolate in time is solely attributed to the initial conditions channels, i.e., our models will predict 40 time-steps from the initial condition time. It is important to emphasize again that the training samples include only 6 wells (4 producers and 2 injectors) while testing samples include any number of wells from a single well up to 9 wells and 3 wells up to 12 wells for the single-phase and two-phase models respectively. This is done to challenge the models to the very limits, because adding/reducing wells alters the pressure and saturation schemes significantly. And since the models are able to predict accurately on the test samples, this proves that the models managed to fully learn the physics not partially. Fig \ref{fig:stats} shows the training, validation and test statistics of the single-phase and two-phase models. 


\begin{figure}
    \centering
    \includegraphics[scale=0.8]{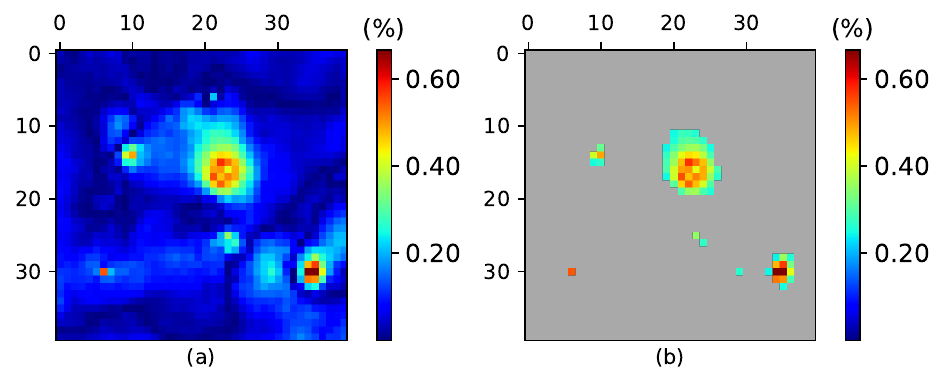}
    \caption{(a) - relative error colormap of a single sample at a single timestep. (b) grid-blocks with highest 5\% relative error.}
    \label{fig:errormask}
\end{figure}

For pressure prediction error analysis we chose to work with the maximum mean relative error MMRE, which is the temporal-maximum highest 5\% spatial-mean relative error. For more clarification, a pressure prediction shape is $(N_x,N_y,N_t)$. We take the mean of the highest 5\% grid-blocks with highest relative errors for all snapshots $N_t$, resulting in an array of length $N_t$, then we take the maximum error of the $N_t$ errors array. Example of the highest 5\% error is shown in Fig.\ref{fig:errormask}, and MMRE is defined as in eq. \ref{MMRE} below

\begin{align}\label{MMRE}
    MMRE=max \biggl\{ \bar{E}_1, \bar{E}_2, \cdots, \bar{E}_{N_t} \biggr\}
\end{align}
$\bar{E_t}$ is the mean relative error of the highest 5\% at timestep $t$ and is given by the following equation.
\begin{align}
    \bar{E_t}=\dfrac{1}{N_{h5}}\sum\limits_{i=1}^{N_{h5}} \bigg| \dfrac{y_{pred}^{(i,t)}-y_{sim}^{(i,t)}}{y_{sim}^{(i,t)}} \bigg|
\end{align}

where $N_{h5}$ is the number of grid-blocks with the highest 5\% error, $y_{pred}$ is model prediction, and $y_{sim}$ is simulation prediction. The reason we chose to work with MMRE as a performance metric and not simply the mean relative error MRE or the maximum relative error is that the maximum relative error occurs at a single grid-block and it can be too conservative because the error at a single grid-block can not represent the overall performance of the model over the entire reservoir. On the other hand, MRE is too lenient since in low permeability areas pressure changes between time-steps are very low and often close to zero, and thereby MRE over the entire spatial domain underestimates performance.

\begin{figure}
    \centering
    \includegraphics[scale=0.5]{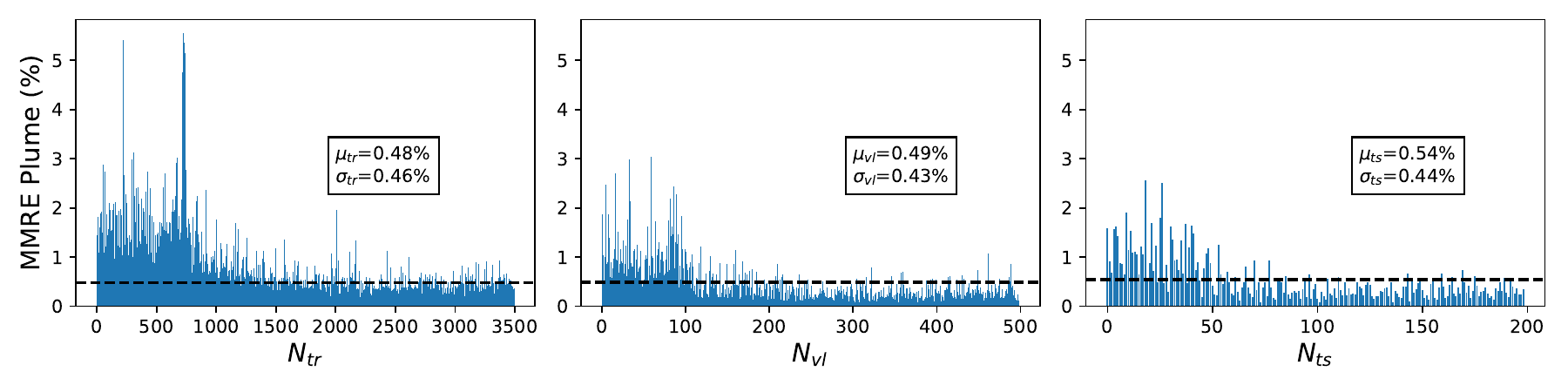}
    \caption{MMRE of plume saturation for training, validation, and testing sets of the two-phase model. $\mu$, $\sigma$ are the mean and standard deviation respectively. Notice the relatively high errors for the first 20\% of each set. The dashlines are the mean $\mu$.}
    \label{fig:plumeMRE}
\end{figure}

Similar to pressures, for saturations we chose to work with maximum-mean-relative error MMRE of the water plume only; excluding grid-blocks where the water plume hasn't reached. Fig. \ref{fig:plumeMRE} displays MMRE of saturations. Note that the first 20\% of the training, validation, and testing samples exhibit higher MMRE than the rest, this is due to the fact that the first 20\% samples are manually assigned initial conditions to be at $t=0$ days. At early times, the volume of water plume evolve from zero, and plume front is much faster than later times, thus, higher errors for saturation predictions are associated with earlier times. For that reason, during training we manually assign 10\%-20\% of the samples to have initial conditions to be at $t=0$ in attempt to force learning at early times.

It is important to note that the time to simulate 100 samples takes about 50 and 117 minutes using CMG reservoir simulator where we use a constant fine timestep size of $0.1$ days and $0.5$ days for single-phase and two-phase respectively, while the prediction of 100 samples using both models takes between 3 and 7 seconds respectively, and thereby achieving acceleration of $\sim$ x1000. For larger reservoir models the acceleration factor would be between 4-6 orders of magnitude higher. Training was performed on NVIDIA A100-PCIE-40GB.

\begin{figure}
    \centering
    \includegraphics[scale=0.65]{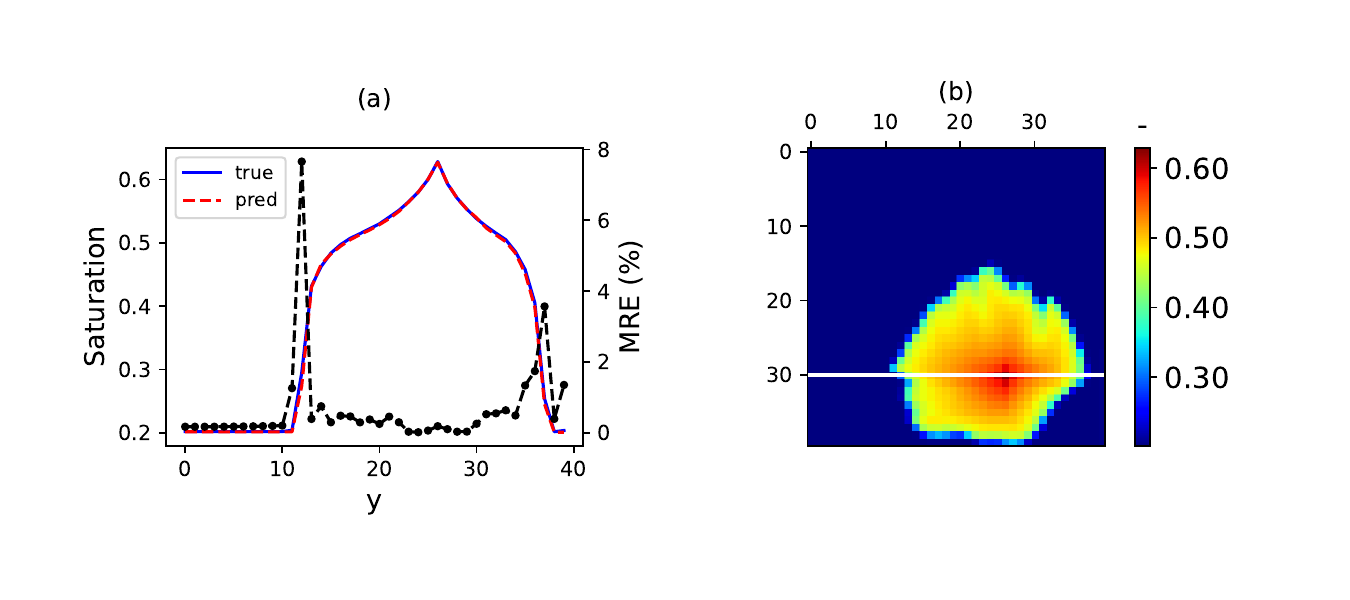}
    \caption{(a) - Saturation true and prediction plot at x-cut=30, the dashed black line is MRE. Notice the spike in MRE at the saturation front. (b) - saturation image plot of the same test sample. The white line is x-cut=30.}
    \label{fig:sat_error}
\end{figure}

\begin{table}
\begin{center}
\caption{FNO architecture}
\label{fno_table}
\small
\begin{tabular}{| l | l| l | l |}
    \hline
    \text{Part} & \text{Layer} & \text{Model 1 Output Shape} & \text{Model 2 Output Shape}\\
    \hline
     & - & - & - \\
    Input& - & (4, 40, 40, 40) & (6, 40, 40, 40) \\
    Lifting& Linear & (36, 40, 40, 40) & (36, 40, 40, 40) \\
    Fourier 1& Fourier3d / Conv1d / Add / ReLU & (36, 40, 40, 40) & (36, 40, 40, 40) \\
    Fourier 2& Fourier3d / Conv1d / Add / ReLU & (36, 40, 40, 40) & (36, 40, 40, 40) \\
    Fourier 3& Fourier3d / Conv1d / Add / ReLU & (36, 40, 40, 40) & (36, 40, 40, 40) \\
    Fourier 4& Fourier3d / Conv1d / Add / ReLU & (36, 40, 40, 40) & (36, 40, 40, 40) \\
    Fourier 5& Fourier3d / Conv1d / Add / ReLU & (36, 40, 40, 40) & (36, 40, 40, 40) \\
    Fourier 6& Fourier3d / Conv1d / Add / ReLU & (36, 40, 40, 40) & (36, 40, 40, 40) \\
    U-Fourier 1& Fourier3d / Conv1d / UNet3d / Add / ReLU & (36, 40, 40, 40) & (36, 40, 40, 40) \\
    U-Fourier 2& Fourier3d / Conv1d / UNet3d / Add / ReLU & (36, 40, 40, 40) & (36, 40, 40, 40) \\
    U-Fourier 3& Fourier3d / Conv1d / UNet3d / Add / ReLU & (36, 40, 40, 40) & (36, 40, 40, 40) \\
    Projection 1& Linear / ReLU & (128, 40, 40, 40) & (128, 40, 40, 40) \\
    Projection 2& Linear & (1, 40, 40, 40) & (2, 40, 40, 40) \\
    \hline
\end{tabular}
\end{center}
\end{table}

\begin{table}
\begin{center}
\caption{Physical Properties.}
\label{physical_prop}
\begin{tabular}{| l | l | l |}
    \hline
    Property & Model 1 & Model 2\\
    \hline
     & - & -   \\
    Porosity; $\phi$ & 0.10 & 0.10 \\
    Viscosity; $\mu$ & $2$ $[cp]$ & $2$ $[cp]$ \\
    Fluid compressibility; $c_f$ & $5.0^{-8}$ $[Pa^{-1}]$ & $5.0^{-8}$ $[Pa^{-1}]$  \\
    Total compressibility; $c_t$ & $5.0^{-8}$ $[Pa^{-1}]$ & $5.0^{-8}$ $[Pa^{-1}]$ \\
    Wellbore radius; $r_w$ & $0.1$ $[m]$ & $0.1$ $[m]$ \\
    Skin factor; $s$ & $0.0$ & $0.0$ \\
    Pressure INCON; $p_0$ & $30$ $[MPa]$ & $30$ $[MPa]$ \\
    Saturation INCON; $S_{w0}$ & - & $0.2$ \\
    Total X Length & $280$ $[m]$  & $280$ $[m]$ \\
    Total Y Length & $280$ $[m]$ & $280$ $[m]$ \\
    Total Z Length & $5$ $[m]$ & $5$ $[m]$ \\
    Total time range & $[0,440]$  $days$ & $[0,3650]$  $days$ \\
    \hline
\end{tabular}
\end{center}
\end{table}

\begin{figure}
    \centering
    \includegraphics[scale=0.7]{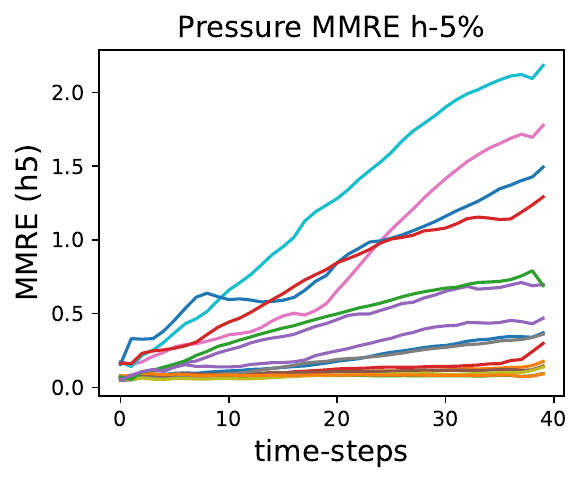}
    \caption{Pressure MMRE of the highest 5\% grid-blocks vs time-steps for 15 random testing samples for the single-phase model. Notice the average increasing trend.}
    \label{fig:growingtrend_1ph}
\end{figure}

\begin{figure}
    \centering
    \includegraphics[scale=0.8]{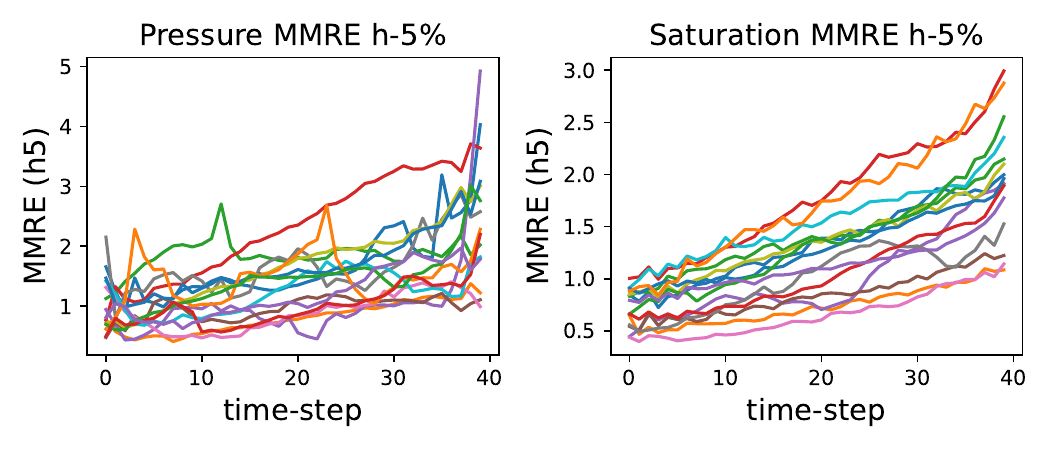}
    \caption{Pressure and saturation MMRE of the highest 5\% grid-blocks vs time-steps for 15 random testing samples for the two-phase model. Notice the average increasing trend.}
    \label{fig:growingtrend_2ph}
\end{figure}

\section{Summary and Discussion}
In this section we discuss the results and some insights that we acquired from this work. In Fig. \ref{fig:stats}, we can see that pressure MMRE of the two-phase model are higher than pressure MMRE of the single-phase model. This initially led us to believe that maybe splitting the two-phase model into two independent models, one for pressures and one for saturations might reduce MMRE, however this didn't show any advantage. Pressures and saturations are coupled and physically speaking can not be decoupled from each other. Pressure wave travels much faster than saturation plume, intuitively one can think of the pressure wave as the leader that travels first and guides the evolution of the saturation plume, this logic is applied in the famous implicit-pressure explicit-saturation (IMPES) which is widely used in traditional reservoir simulations. For that reason, we insist on exploring single-model architectures that predict both pressure and saturation simultaneously.

In Fig. \ref{fig:2ph_sw_pred} it is evident from the absolute error plots (5th column) that the highest errors in saturation predictions are on the front of the plume. This is better illustrated in Fig. \ref{fig:sat_error} that shows the saturation MRE of a test sample at a specific time-step where it is clear that MRE spikes at the front. The front of the plume is associated with the highest gradients $dS_w/dx$ and $dS_w/dy$. Initially we thought that since the solution is being learnt in Fourier space, the high errors at the front might be caused by Gibbs ringing phenomena at step functions similar to the saturation front \cite{gibbs_gottlieb, Hewitt1979TheGP}. For that reason we took higher number of Fourier modes $m=20$, however this didn't enhance the results. We also tried masking schemes that put more attention to grid-blocks associated with highest gradients, but that didn't produce no worth-mentioning advantage. Classically, reducing errors at the front requires finer mesh which leads to greater training and simulation time.

Although this work is not mainly centered around minimizing errors to the ideal, several masking and attention mechanisms were utilized in attempt to reduce errors in the two-phase model to be similar to those in the single-phase model with no worth-mentioning success.

What we are able to verify is that the complexity of the model has a major role on errors, for example, holding well controls constant for all samples reduces errors.




Another observation we have noticed, is that errors increase with time-steps. Remember, the model predicts 40 consecutive time-steps simultaneously. Fig. \ref{fig:growingtrend_1ph} and Fig. \ref{fig:growingtrend_2ph} illustrate the relationship between MMRE of the highest 5\% grid-blocks and the number of time-steps where it is evident that on average later time-steps are associated with higher errors. In Fig. \ref{fig:growingtrend_2ph} the trend is steadier for saturation predictions than pressure predictions. Consequently, to be more conservative with errors one can simply train a model with a reduced timestep prediction range. In this case for example, if we reduce the timestep prediction range to 20 (instead of 40), the input and output shapes would be $(N_s,N_c,40,40,20)$ and $(N_s,N_o,40,40,20)$ respectively, where $N_o$ is the number of output channels; 1 for single-phase model and 2 for two-phase model.

Such models can work in conjunction with traditional simulators to further reduce errors. The prediction of the model can be fed into a simulator as an initial guess for a specific timestep, and inside the simulator we can perform couple of Newton-Raphson iterations until convergence. We don't see the significance and value of this model in the relatively small errors only but mostly in its ability to predict on completely unseen scenarios regarding permeability fields and well specifications.


\section{Conclusions}
\label{sec:conclusions}
This work presents a single Fourier neural operator based model that can accurately predict pressure and saturation distributions for unseen permeability fields, well locations, number of wells, well controls as well as extrapolation in time. We achieved this by constructing the input tensor in a binary fashion which contains both well locations and well controls values. Additionally, the data augmentation technique we employed enriched our dataset and prevented overfitting. It also reduced the number of simulations by 75\% . The significance of this work is that it is able to learn the complete physics from data. After the model is trained, predictions become available within seconds, and therefore, the ability to predict accurate results on new permeability fields significantly accelerates history matching. Likewise, the ability to predict on new well locations, well controls, and number of wells is particularly useful for hydrocarbon as well as carbon storage reservoir optimization and future planning of new wells, thereby allowing reservoir engineers to quickly identify the optimum locations to place new wells and to quickly understand the consequences without running new simulations.

We believe that such models can be a cornerstone in the process of developing reservoir digital twins that enable online interaction between users and reservoir parameters.

\section{Acknowledgements}
Part of this work was funded by Xecta Labs. E. Gildin and D. Badawi gratefully acknowledge Xecta's team; Sathish Sankaran, Zhenyu Guo, Hardik Zalavadia and Prithvi Singh Chauhan for the financial and technical support throughout this work.

\section{Code  Availability}
The source code for this work and the datasets are publicly available in the link: \url{https://github.com/daniel-1388/twophase-neuraloperator}. 

 \bibliographystyle{unsrtnat} 
 \bibliography{references.bib}
\end{document}